# A non-invasive method for nanoscale electrostatic gating of pristine materials

Arjan J.A. Beukman[1*], Fanming Qu[1*], Ken W. West[2], Loren N. Pfeiffer[2], Leo P. Kouwenhoven[1†]

[1] QuTech and Kavli Institute of Nanoscience, Delft University of Technology, GA 2600 Delft, The Netherlands.
[2] Department of Electrical Engineering, Princeton University, Princeton, New Jersey 08544, USA
[†] Email：L.P.Kouwenhoven@tudelft.nl
*Equal contribution.

**Electrostatic gating is essential for defining and control of semiconducting devices. However, nano-fabrication processes required for depositing gates inevitably degrade the pristine quality of the material of interest. Examples of materials that suffer from such degradation include ultra-high mobility GaAs/AlGaAs two-dimensional electron gases (2DEGs), graphene, topological insulators, and nanowires. To preserve the pristine material properties, we have developed a flip-chip setup where gates are separated from the material by a vacuum, which allows nanoscale electrostatic gating of the material without exposing it to invasive nano-processing. An additional benefit is the vacuum between gates and material, which, unlike gate dielectrics, is free from charge traps. We demonstrate the operation and feasibility of the flip-chip setup by achieving quantum interference at integer quantum Hall states in a Fabry-Pérot interferometer based on a GaAs/AlGaAs 2DEG. Our results pave the way for the study of exotic phenomena including fragile fractional quantum Hall states by preserving the high quality of the material.**

Electrostatic gating is widely used to define functional electronic devices on semiconductors[1]. Various growth techniques persistently produce increasingly high-quality semiconducting materials that form the basis of a wide variety of devices. But degradation of the pristine material quality by nano-fabrication processes hinders exploration of the devices' full potential[2, 3]. One particular example is ultra-high mobility two-dimensional electron gases (2DEGs) in GaAs/AlGaAs quantum wells with a carrier mobility exceeding $3.5\times10^7$ cm$^2$/Vs [4,5]. Such high-quality material enables the study of novel interaction phenomena including the fractional quantum Hall (FQH) effect, where strong electron-electron interactions provide a wealth of new phases[6-11]. In particular, the 5/2 and 12/5 FQH states are believed to obey non-Abelian quantum statistics[12-15]. Attempts to discover the non-Abelian nature of the 5/2 state have been guided by promising proposals for a Fabry-Pérot interferometer (FPI) defined by electrostatic gating[16,17]. Quantum interference has been achieved at integer quantum Hall (IQH) states[18-21] as well as at some FQH states[22,23], but remains a challenge when applied to the fragile 5/2 state[24-26]. Depending on the electrostatics, the mechanism for the interference can be of Aharonov-Bohm (AB) type or Coulomb-dominated (CD)[19,20,27-29]. So far, all devices for use in such experiments have been conventionally fabricated by depositing metal on the surface of the ultra-high-mobility GaAs/AlGaAs using electron-beam lithography. However, it is widely accepted that nano-fabrication processes degrade the quality of the pristine 2DEG and thus weaken the fragile states[2,10,11,30-32], preventing exploration of the underlying physics. Therefore, it is crucial to maintain the as-grown material quality while inducing the electrostatic confining potential required for interferometry.

The present research takes a new approach that preserves the pristine quality of the heterostructure material while enabling nanoscale gating. The gate structure is fabricated on a separate chip that is flipped over and brought close (~100 nm) to the heterostructure of interest, resembling a flip-chip assembly[33,34]. FPIs are fabricated on the gate-chip and assembled onto a high mobility GaAs/AlGaAs chip. At low temperatures and in high magnetic fields stable quantum interference patterns at IQH states are measured, an essential step towards interferometry at the FQH states.

The flip-chip technique has several unique advantages compared to the conventional gating method[34]. First, all the invasive nano-fabrication processes are performed on a separate chip. Therefore the GaAs chip is not contaminated with resist residues; no damage is inflicted on the surface; the device region under study is not exposed to any electron-beam radiation[2,31], etc. Second, the vacuum gap between the gate and the surface of the GaAs chip is an ideal dielectric layer, solving the problem of gate-leakage, interface charges, defects in the dielectric and hysteresis in gating that are often found in traditional gate-dielectrics[11, 32, 35]. Third, the GaAs chip and the gate-chip are easily assembled and disassembled, and both can be reused. Fourth, the flip-chip setup enables complex and flexible device designs as the 'dirty' nano-processing is performed on a separate chip. One can think of multi-layered gate/dielectric structures for various functions, e.g., single electron transistor sensors and superconducting quantum interference devices built into the gate-chip. Moreover, the setup can also be applied to other materials sensitive to nano-fabrication processes[3, 36], for instance, graphene, carbon nanotubes, HgTe/HgCdTe quantum wells and three-dimensional topological insulators.

Several constraints have to be considered in the design of such a flip-chip setup: First, to electrostatically confine the potential landscape in the 2DEG on a length scale below 100 nm, the gates need to hover 100 nm or less above the surface of the target chip. Second, to prevent significant charge noise, the relative vibration amplitude between the two chips should be sub-angstrom. Third, the setup must sustain the cooling from room temperature to millikelvin temperatures as well as operation in high magnetic fields. Fourth, due to the close separation over a large area of ~3 mm$^2$ the two chips need to be absolutely parallel (angle << 1 mrad). Fifth, a precise positioning of the gate pattern with respect to the target chip is required, e.g. positioning gates over a graphene flake. Finally, the setup should allow for illumination of the 2DEG.

Figure 1a schematically depicts the design of our flip-chip assembly. A double-side polished quartz chip (blue) supports the Ti/Au (5 nm/40 nm) gates (yellow). Titanium posts (black) with a thickness of 100 nm are evaporated onto the quartz chip. When the quartz chip is flipped over and pressed against the target chip, these taller posts ensure a fixed height of the gates with respect to the surface. Furthermore, the direct contact of the Ti posts with the target substrate results in a good mechanical stability.

Although all the aforementioned fabrication processes are performed in a cleanroom, irregularities such as dust particles that prevent the two chips from approaching each other can still be present on the ~3 mm$^2$ surfaces. The quartz surrounding the gates and posts is etched ~4 µm inward by dry reactive ion etching to reduce the area that is in close proximity to the target chip. All the structures on the gate chip are patterned by standard electron-beam lithography. The transparent material of the gate-chip allows for precise alignment of the gates relative to the target chip simply by optical means. Moreover, it enables illumination of the 2DEG through the gate-chip. The insulating nature of the quartz substrate requires several precautionary measures: First, to prevent charge accumulation during the electron-beam exposure, a 15 nm thick Cr conductive layer is sputtered onto the quartz chip prior to spin coating of PMMA resist and is subsequently etched away after development. Second, as the insulating substrate is prone to static electricity, the fine gates are easily damaged and dust particles can be attracted onto the surface. Therefore, throughout the fabrication process, additional metal bridges are used to interconnect all the gates to impose an equal potential on all gates. Once the setup assembly is complete, the bridges are scratched away.

In the present study we used a high mobility GaAs/AlGaAs heterostructure as the target material, where the 2DEG is 100 nm deep with single-side Si doping. To minimize the amount of fabrication on the heterostructure, six ohmic contacts are formed by soldering indium droplets on the edges of the heterostructure followed by a rapid thermal annealing step. Occasionally the surface roughness (~100 nm on a ~mm scale) of the target wafer prevents a uniform spacing between the two chips. Regions with a larger separation require more negative gate voltage to deplete the electrons underneath the gate, resulting in gate leakage at regions with small separation. To avoid such gate leakage we performed one electron-beam lithography and wet etching step on the GaAs heterostructure to define a narrow (60 µm) transport channel, which is then isolated from the regions with possible gate leakage. For future applications this step needs to be avoided by ensuring the surface roughness of the wafers is much smaller than 100 nm over 3 mm$^2$.

Figure 1b shows a scanning electron microscope picture of a typical gate-chip. The inset shows the interferometer gate pattern for assembly A. The gate-chip and the GaAs chip are assembled and mounted on the cold finger of the dilution refrigerator, as shown in Fig. 1c (see Supporting Information).

Flip-chip assembly A uses a gate-chip with an interferometer gate pattern depicted in the inset of Fig. 1b. Figure 2a shows a schematic of the assembly numbering contacts and gates. The interferometer consists of two quantum point contacts (QPCs) formed by the left two (7 and 10) and right two (9 and 12) gates, and a pair of plunger gates (8 and 11). The QPCs bring the edge channels on the opposite edges in the quantum Hall regime together so that inter-edge scattering becomes possible. The plunger gates control the area of the interferometer and thus the number of electrons. The assembly is cooled down in a wet dilution refrigerator hanging on a passive vibration isolation frame. The still pumping line is modified to isolate the sample from vibrations of the environment (e.g. pumps) but without extra damping stages inside the fridge.

At a base temperature of ~100 mK the 2DEG is first examined without energizing the gates (grounded). Standard low frequency lock-in techniques are used to measure the longitudinal ($R_{xx}=R_{3,4}=(V_3-V_4)/I_{2-5}$) and Hall resistances ($R_{xy}=R_{3,1}=(V_3-V_1)/I_{2-5}$) employing a current bias of 2 nA between contacts 2 and 5 ($I_{2-5}$) in perpendicular magnetic field ($B$), as shown in Fig. 2b. A sheet density of $1.35\times10^{11}$ cm$^{-2}$ and a mobility of $3.3\times10^6$ cm$^2$/Vs are extracted.

Next the QPCs are characterized. As Fig. 2c shows, the conductance $G=I_{2-5}/(V_3-V_4)$ drops when the left QPC is energized with negative voltages ($V_{7,10}$) because the electron density underneath the gates is reduced. At a voltage of $V_{7,10}$=-2.2 V the 2DEG underneath the gates is depleted (see Supporting Information Figure S1). Using the depletion voltage as an input to a capacitor model we estimate the distance between the gate and the surface of the GaAs chip to be ~84 nm. In an independent test of a flip-chip with two quartz chips containing capacitor plates, a separation of ~120 nm is confirmed by measuring the capacitance at high frequencies (see Supporting Information Figure S2). As more negative voltages are applied to the gates, the QPC is pinched off gradually. Note that almost no hysteresis is observed in opposite sweeping directions of the gate voltages, illustrating the advantage of using a vacuum as a gate dielectric. A close-up of the QPC trace in the depleted regime, as Fig. 2d illustrates, shows quantized conductance plateaux due to depopulation of 1D subbands. The right QPC presents an identical behavior (see Supporting Information Figure S3). To investigate its reproducibility, the assembly is cooled down five times in two different cryostats. Depletion and pinch-off voltages do not change (< ±0.1 V) after each thermal cycle, which demonstrates the robustness of the mechanical structure.

To study the vibrational stability of the setup we define a quantum dot in the 2DEG. The voltages of the plunger gates are set past the depletion point ($V_{8,11}$=-2.5 V) and both QPCs are close to pinch-off ($V_{9,12}$=-3.27 V and $V_{7,10}$=-3.32 V). A voltage bias $V_b$ is applied to contact 2, and current is measured between contact 5 and ground. Figure 3a shows the current as a function of $V_b$ and gate voltage $V_8$=-2.5 V+$\Delta V_8$. Regular and stable Coulomb diamonds are observed. A charging energy of ~55 μeV is extracted from the bias voltage at the top of the diamonds. Because a quantum dot is very sensitive to induced charges and hence to changes in capacitance, it is a perfect tool for investigating the stability of the gate-heterostructure separation. Vibrations are expected to shift the Coulomb diamonds up and down in gate-voltage and also change their size. Large vibrations induce extra electrons in the quantum dot and would distort the Coulomb diamonds beyond recognition. Small vibrations result in blurring of the edges of the Coulomb diamonds and broadening of the zero-bias crossing. Therefore, an upper bound on the vibration amplitude can be estimated from the blur at the zero-bias crossing of the diamonds. Using 0.15 mV as the uncertainty of the Coulomb resonance peak position and taking into account the length ratio (~1/3) between gate 8 and all six gates, the upper bound on the vibration amplitude is estimated to be 1.5 pm (see Supporting Information Figure S1). Note that as the Ti posts are more than 50 μm away from the quantum dot, the influence of the floating potential is negligible. The duration of the measurement of the Coulomb diamonds of Fig. 3a is ~15 minutes. On longer timescales, though (~hour), a drift of the Coulomb diamonds is observed (see Supporting Information Figure S4). However, a slow drift such as this is easily corrected using a feedback loop on the gate voltages (not yet implemented). Nonetheless, as discussed later, the current setup is stable enough to perform quantum interference measurement in a time scale of ~30 minutes.

Having confirmed the stability of the setup, we continue to investigate the quantum interference at IQH states using the flip-chip setup. We first explore the response of the QPCs in a perpendicular magnetic field. $V_{8,11}$ are set past depletion to -2.2 V. The magnetic field is set to 1.3 T, where the bulk of the 2DEG is in the incompressible state with filling factor ν=4. A current of $I_{2-5}$=300 pA is applied, which is small enough to avoid self-heating. Diagonal conductance $G_d=I_{2-5}/(V_3-V_6)$ is probed as a direct measure of the filling factor at the QPC, i.e., the number of channels transmitted by the QPC. As each of the two QPCs is energized with negative voltages, quantized conductance plateaux of $f_0 e^2/h$ ($f_0$=1,2,3,4) are observed (see Supporting Information Figure S5), verifying the functioning of the QPCs and identifying the regions where backscattering occurs. Note that $f_0$ here represents the number of fully transmitted chiral quantum Hall edge-modes, unlike 1D subbands for a QPC without magnetic field, as Fig. 2d depicts.

For both QPCs, $G_d$=2$e^2$/h at the depletion voltage of -2.2 V, indicating that the density in the QPCs is around half of that in the bulk (where ν=4). This is due to the smooth potential landscape induced by the gates. In pursuance of an interferometer, gate voltages for the QPCs are set to the low voltage side of plateau 3 ($V_{7,10}$=-1.9 V and $V_{9,12}$=-1.6 V) such that the two outer edge channels are fully transmitted and the third (inner) edge channel is partially transmitted, while the fourth channel is fully reflected. Note that although the QPC voltages are not yet beyond the depletion voltage of -2.2 V, the backscattering observed in the third edge channel indicates that the edge channels are brought close to each other. At plunger gate voltages $V_{8,11}$ = -3.5 V the amplitude of the quantum interference oscillations is maximized.

Figure 3b shows the dependence of $G_d$ on $B$ at the above gate voltage settings. Conductance plateaux of $f_T e^2/h$ ($f_T$=1,2,3,..) are observed, where $f_T$ equals the number of edge-modes that are fully transmitted through the interferometer. At the low magnetic field side of the fully developed plateaux, the innermost edge modes on opposite sides of the QPCs couple, leading to finite inter-edge tunneling. An electron in the innermost edge-channel entering the interferometer is partially reflected at the first QPC. The transmitted electron wave is again partially reflected at the second QPC. The two backscattered electron waves interfere constructively or destructively depending on the flux through the area between the QPCs. As Figs 3c and 3d show, quantum interference oscillations of $G_d$ are observed between plateaux 2 and 3 ($f_T$=2), and 1 and 2 ($f_T$=1), with magnetic field periods of 0.275 mT and 0.547 mT, respectively. From the latter magnetic field period the estimated enclosed area by the edge-modes is 7.6 μm², close to the design area of 10 μm² (as discussed below).

Depending on sample geometry and size, quantum oscillations in a FPI fall into one of two regimes: the Aharonov-Bohm (AB) regime, or the Coulomb-dominated (CD) regime[19, 20, 27-29]. To investigate the effect of the interferometer size we switch to flip-chip assembly B containing a larger interferometer. The GaAs chip used for assembly B is from the same wafer as in assembly A. The QPCs have an opening of 1.7 μm and the lithographic area enclosed by the interferometer is 13.6 μm$^2$ (see inset of Fig. 4a). The electrons underneath the gates are depleted at a gate voltage of -3 V, implying a gate-heterostructure separation of 115 nm (see Supporting Information Figure S6). Despite their large opening, the QPCs can still reach full pinch-off at a voltage of -9 V without hysteresis (see Supporting Information Figure S6).

The build-up of the interferometer follows the same procedure as for assembly A. To observe oscillations with significant amplitude in $G_d$, all gate voltages are set to -4.5 V. As shown in Fig. 4a, fully developed quantized plateaux in $G_d$ are visible over a large magnetic field range. Between the plateaux, oscillations are again observed due to inter-edge tunneling events. Figures 4b, d, and f zoom in on the regions between plateaux 3-4 ($f_T$=3), 2-3 ($f_T$=2) and 1-2 ($f_T$=1), and show interference oscillations as a function of magnetic field with periods of 0.22 mT, 0.29 mT, and 0.56 mT, respectively. In addition to oscillations in the magnetic field, we observe oscillations as a function of plunger gate voltage between plateaux 3-4, 2-3, and 1-2 as depicted in Figs. 4c, e, and g, respectively. The periods are 2.4 mV, 2.02 mV and 2.11 mV respectively and exhibit no obvious dependence on 1/B (as discussed below and Supplementary Figure S7).

Furthermore, the interference pattern can be mapped out as a function of both perpendicular magnetic field and plunger gate voltage, as shown in Figs. 5a, b, and c for the regions between plateaux 3-4, 2-3, and 1-2 respectively. In all the measured 2D maps, the parallel lines of constant phase disperse with a positive slope (as discussed below). Each plot is obtained in ~35 minutes, indicating a very stable operation of the flip-chip setup with negligible charge noise, consistent with the quantum dot characterization.

In a FPI, an electron wave in the inner edge state partially scatters at each QPC to the opposite backward-flowing edge channel. These back-scattered electron waves interfere with each other with a relative phase set by the flux enclosed by the trajectories, $\Phi=BA$ (where $B$ is the magnetic field, and *A* the area of the interferometer) i.e., the Aharonov-Bohm phase. The period of the interference in *B* is $\Delta B=\varphi_0/A$ ($\varphi_0=h/e$) regardless of the filling factors at the QPCs. On the other hand, the period in gate voltage is $\Delta V_g=\alpha\, \varphi_0/B$, where α represents how fast the area changes upon changing the gate voltage. However, this single-particle picture breaks down when Coulomb interactions are strong. The interior of the FPI forms an island similar to a quantum dot with a finite charging energy[27, 37]. Depending on the electron-electron interaction strength and the capacitances from the island to the gates and the edge channels, the FPI can enter the CD regime. The charging energy of the island is given by[19, 27]:

$$E = \frac{1}{2C}\left(eN + ef_T BA/\phi_0 - C_g V_g\right)^2, \tag{1}$$

where *N* is the number of electrons on the island, *C* and *A* are the total capacitance and the area of the island, respectively, and $f_T$ the number of fully occupied (transmitted) Landau levels. $C_g$ and $V_g$ are gate-to-island capacitance and gate voltage, respectively. Magnetic field couples to the charge on the island through the $f_T$ underlying Landau levels. The island can be charged with one electron by changing the flux through the island by $1/f_T \cdot \varphi_0$ or the gate voltage by $e/C_g$. The induced charge on the island alters the area of the interference loop because of Coulomb interactions. Therefore, the magnetic field affects the enclosed flux $\Phi=BA$ through *B* and *A*, in contrast to the AB regime. The period of the oscillations in *B* and $V_g$ are $\Delta B=1/f_T \cdot \varphi_0/A$ and $\Delta V_g=e/C_g$ (as both generate a phase difference of 2π between the two interfering waves), respectively.

In addition, $G_d$ can be plotted as a function of $V_g$ and *B*, as shown in Fig. 5. For the AB case, increasing *B* is equivalent to increasing $V_g$ as they both increase the flux. A negative slope is therefore expected for lines with constant phase . However, for the CD case, increasing *B* is equivalent to reducing $V_g$ (see eq.1), thus leading to a positive slope of the constant phase lines.

In assembly A, the oscillations of $G_d$ follow $\Delta B \propto 1/f_T$, indicating operation in the CD regime. In assembly B, $\Delta B \propto 1/f_T$, and oscillations in $V_g$ show no dependence on $f_T$. In addition, the constant phase lines in the 2D maps (Fig. 5) have a positive slope at all measured plateaux. Thus, assembly B also falls into the CD regime, despite the larger dimension and wider QPCs of the interferometer. The CD type of interference is attributed to the small capacitance of the island, which is dominated by the capacitance to the edges of the gates. A top gate can increase the island's total capacitance and turn the FPI into the AB regime[19, 20]. Moreover, a top gate would generate a sharper potential landscape (see Supporting Information Figure S8) and thus a better control over the interferometer.

We observe a small drift of the conductance peaks in the quantum dot, which we attribute to slow relaxation of the material and/or the setup. This slow drift can be actively corrected by a feedback loop on the gate voltages. Nevertheless, the current setup is stable enough to study the physics of the IQH and FQH states. In this flip-chip design, the gate-chip and material chip are mechanically connected by the posts. The influence of this mechanical contact on the material properties needs further study. The yield of flip-chip assemblies that meets all requirements is mostly determined by the flatness of both chips. Substrates without mechanical deformations would increase the yield, and eliminate the need to etch channels in the target chip.

We have developed a flip-chip setup that allows for nanoscale electrostatic gating without the requirement of extensive nano-fabrication processing on the target material. By measuring quantized conductance in a QPC and Coulomb blockade features in a quantum dot we have shown that the flip-chip setup operates as required, with a gate-heterostructure separation of ~100 nm and vibration amplitudes of several picometers. After characterization of the setup we performed measurements at high magnetic fields. Using a FPI gate-pattern on the gate-chip, we probed the interference of IQH edge states. In both small and large interference loops (assembly A and B, respectively) the quantum interference is dominated by electron-electron interactions, i.e., the Coulomb-dominated regime. To obtain pure Aharonov-Bohm interference the capacitance of the interferometer needs to be increased, achievable by implementing an extra top gate on the gate-chip.

In addition to its application in the high-mobility heterostructures investigated in this work, the flip-chip setup can be used on a wide range of materials, such as flakes of graphene or of 3D topological insulator. The platform developed for non-invasive nanoscale gating presented in this research opens a promising route toward the development of devices of unprecedented quality.

The authors thank M. van der Krogt, M. Zuiddam for help with device fabrication, T.H. Oosterkamp, G.A. Steele, S. Goswami, J.D. Watson and S. Nadj-Perge for valuable discussions and comments. This work has been supported by funding from NWO/FOM and Microsoft Corporation Station Q.

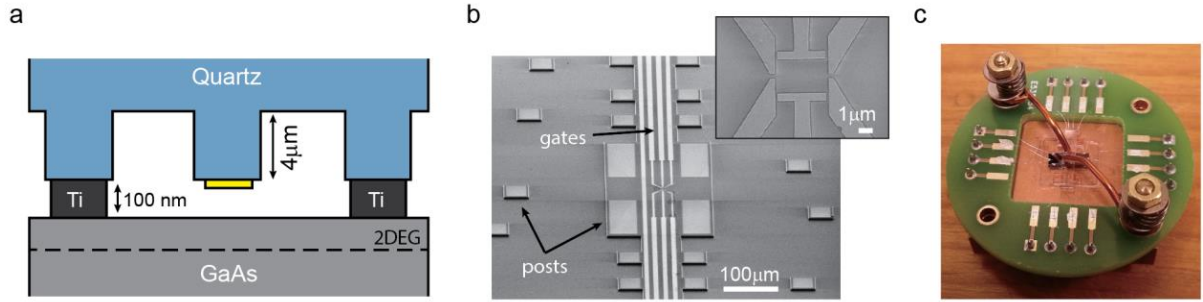

**Figure 1:** The flip-chip setup. (**a**) Schematic of the flip-chip setup (not to scale). The Ti/Au (5 nm/40 nm) metallic gates (yellow) and Ti posts (100 nm, black) are deposited on a quartz chip (blue). The quartz surrounding the gates and posts is etched inward by ~4 μm. Afterwards, the quartz chip is flipped over and put onto the GaAs target chip (grey), which has not underwent invasive fabrication. The posts, which are by design taller than the gates, ensure a minimum gate-heterostructure separation of ~60 nm. The 4 μm deep etch of the quartz chip reduces the probability that dust particles prohibits close separation. (**b**) Scanning electron micrograph of the quartz chip (from assembly A) showing the gate-pattern (inset) and posts. (**c**) Optical image of the cold finger onto which the flip-chip assembly is mounted. Copper springs with adjustable forces press on the quartz chip to ensure a mechanically stable and small separation upon cooling to ~100 mK. The diameter of the PCB board is 3.5 cm.

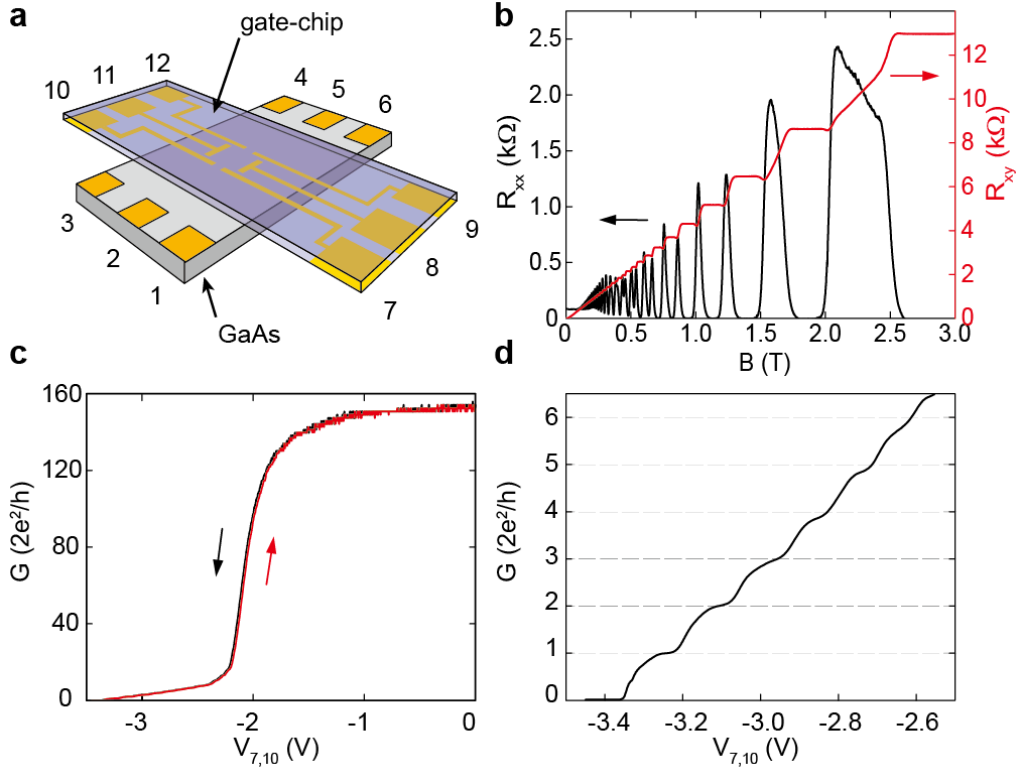

**Figure 2:** Characterization of the 2DEG and gates for assembly A. (**a**) Schematic of the flip-chip assembly with contact and gate numbers. (**b**) Longitudinal (black) and Hall (red) traces as a function of perpendicular magnetic field $B$ with all gates grounded. (**c**) Conductance as a function of gate voltage $V_{7,10}$ on the left QPC (gates 7 and 10) in both sweeping directions, indicating depletion of the 2DEG underneath the gates around $V_{7,10}$=-2.2 V and pinch-off of the QPC around $V_{7,10}$=-3.4 V. (**d**) Zoom-in on **c** near pinch-off showing quantized conductance steps.

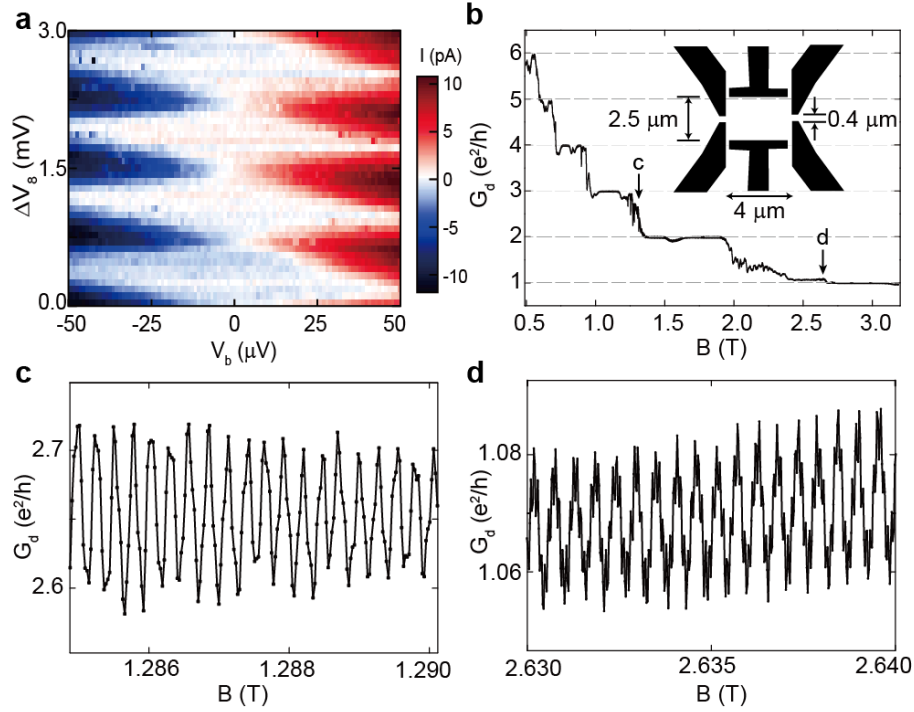

**Figure 3:** Quantum dot and interferometer formed by the gates in assembly A. (**a**) Coulomb diamonds measured on a quantum dot defined by energizing all gates. (**b**) Diagonal conductance $G_d$ as a function of perpendicular magnetic field $B$. The gate voltages are tuned to $V_{8,11}$ =-3.5 V, $V_{7,10}$=-1.6 V, and $V_{9,12}$=-1.9 V. The quantum interference oscillations are plotted in **c** for the region between plateaux 2 and 3, and **d** between plateaux 1 and 2.

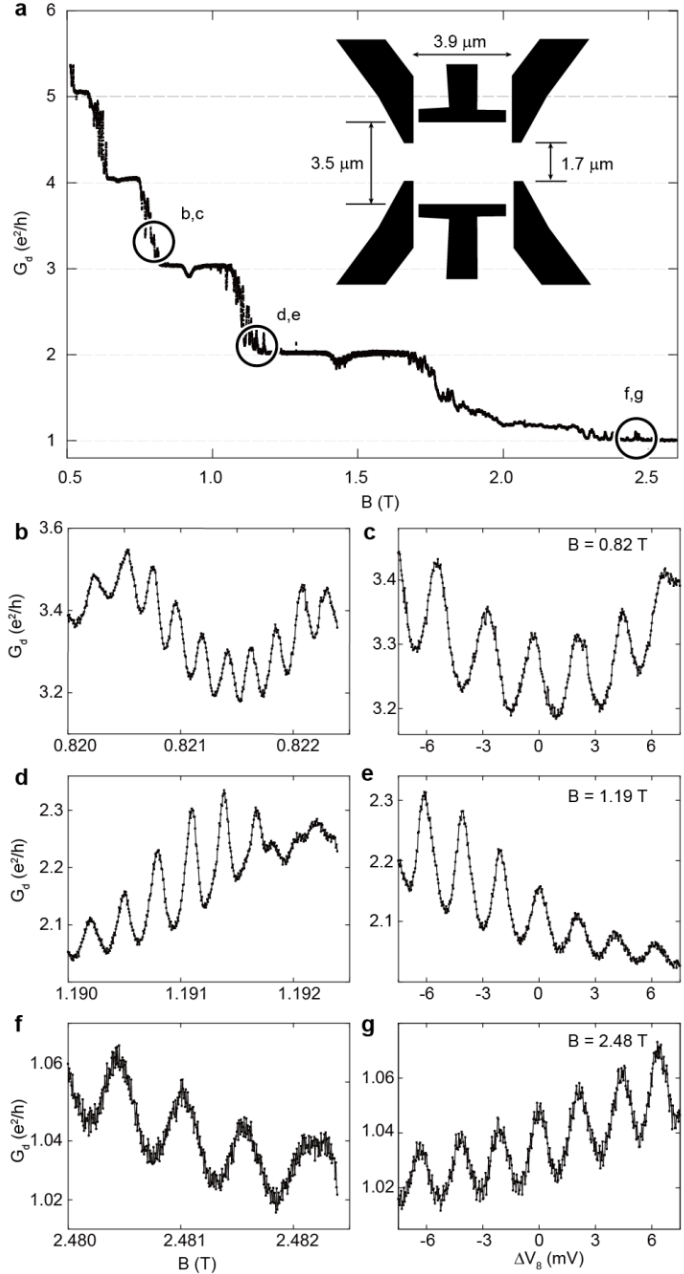

**Figure 4:** Quantum oscillations at IQH states as a function of magnetic field and gate voltage for assembly B. (**a**) $G_d$ as a function of $B$ showing well-quantized integer plateaux. Inset depicts the gate-pattern for assembly B, with a QPC separation of 1.7 μm and an enclosed area of 13 μm². (**b**, **d**, **f**) and (**c**, **e**, **g**) Quantum oscillations as a function of $B$ and plunger gate voltage $V_8$=-4.5 V+$\Delta V_8$ between plateaux 3 and 4, 2 and 3, and 1 and 2, respectively.

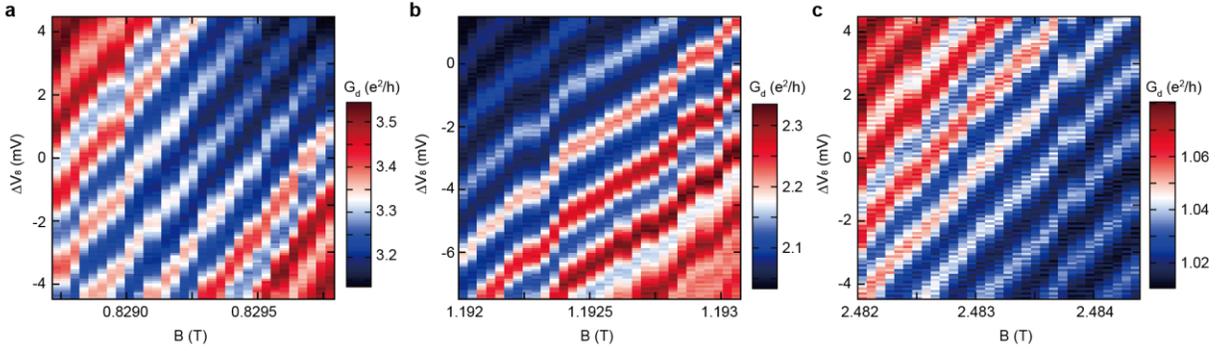

**Figure 5:** Quantum oscillations as a function of both magnetic field and gate voltage for assembly B. (**a**, **b**, **c**) Diagonal conductance $G_d$ as a function of $V_8 = -4.5$ V$+\Delta V_8$ and $B$ for plateaux 3-4, 2-3, and 1-2, respectively. For all three regions the parallel lines with a constant phase disperse with a positive slope.

**Supporting Information**

# A non-invasive method for nanoscale electrostatic gating of pristine materials

Arjan J.A. Beukman[1*], Fanming Qu[1*], Ken W. West[2], Loren N. Pfeiffer[2], Leo P. Kouwenhoven[1†]

[1] QuTech and Kavli Institute of Nanoscience, Delft University of Technology, GA 2600 Delft, The Netherlands.

[2] Department of Electrical Engineering, Princeton University, Princeton, New Jersey 08544, USA

[†] Email：L.P.Kouwenhoven@tudelft.nl

*Equal contribution.

**Methods**

The flip-chip assembly consists of a gate-chip and a material chip, an 8×8 mm quartz piece used as base and additional quartz supports to hold the gate-chip in place. The material chip is first glued onto the quartz base using S1805 photoresist. Then two quartz supports with the same thickness as the material chip are glued on either side of the material chip on the 8×8 mm piece. Afterwards, droplets of photoresist are placed on these quartz supports and the gate-chip is positioned over the material chip and quartz supports in a mask aligner. After alignment, a small force is applied onto the gate-chip to fix the setup while the glue dries (for ~12 h). The flip-chip assembly can then be safely transferred out of the cleanroom without dust entering between the chips. The assembly is glued directly on the cold finger of the cryostat and all contacts are connected to a PCB board through a combination of wire bonding, indium soldering and use of silver paint. Next, copper clamps are placed onto the gate-chip. Springs attached to the cold finger can adjust the force exerted by the copper clamps. The forces and positions of the clamps are fine-tuned such that a uniform dark blue color is observed suggesting close and uniform spacing between the chips. This color originates from the interference between the reflected light from the surface of the material chip and the reflected light from the bottom surface of the gate-chip. A dark blue color suggests a separation of ~100 nm between the gate and the surface of the material chip, and a uniform color signals uniform spacing. Figure 1c shows the flip-chip assembly mounted on the cold finger including the copper springs. The whole setup is free from magnetic parts and fits in a standard 2 inch magnet bore.

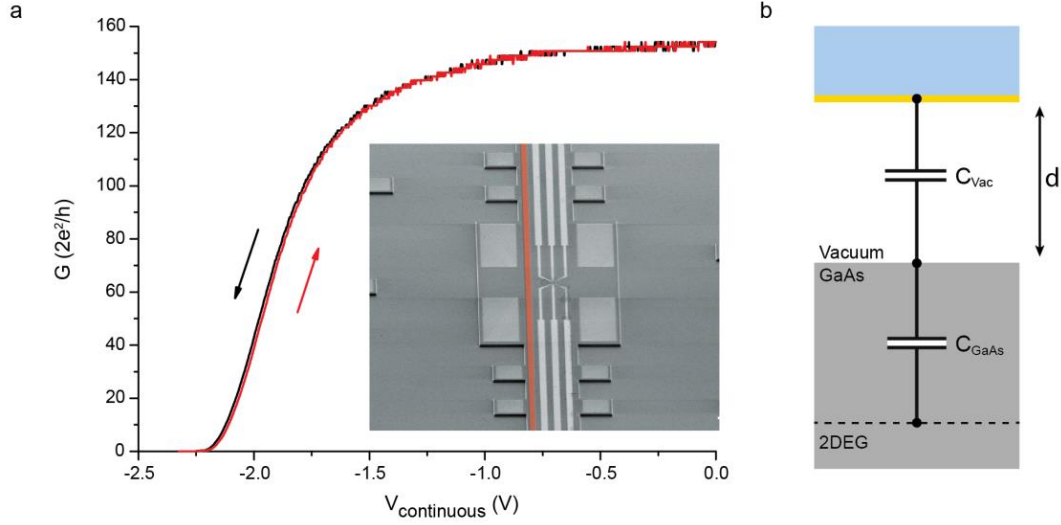

**Figure S1: Depletion voltage for the continuous gate that spans the width of the 2DEG in assembly A in the main text.** (**a**) Four-terminal conductance as a function of the voltage on the continuous gate (indicated by red in the false-colored SEM image in **a**). The gate has a width of 8.3 μm, much wider than the average gate-heterostructure separation of ~100 nm. Therefore, the parallel plate capacitor model (**b**) is justified for estimating the vacuum gap between the two chips. At a voltage of $V_{continuous}$=-2.2 V the conductance drops to zero, indicating that all electrons underneath the gate are depleted. The gate-2DEG coupling can be described by the capacitor network depicted in **b**. $C_{GaAs}$ is the capacitance per unit area between the 2DEG and the surface of the GaAs. The capacitance between the gate and the surface of GaAs is $C_{vac}=\varepsilon_0\varepsilon_r/d$ per unit area. The total capacitance between the 2DEG and the gate is $C_t = 1/(1/C_{vac}+1/C_{GaAs})$.

Using $e\cdot n = C_t V_{continuous}$ and $V_{continuous}$ = -2.2 V, a density of $n$ = 1.35×10$^{11}$ cm$^{-2}$ (obtained from Hall measurements), a depth of the 2DEG of 100 nm, we estimate that $d$ = 84 nm.

The vibration amplitude between the two chips is estimated using the blur at the zero-bias crossings of the Coulomb diamonds. From Fig. 3a in the main text we estimate a 20% blur (0.15 mV in $V_8$). A change in $d$ affects the island potential through all six gates ($V_{8,11}$=-2.5 V, $V_{9,12}$=-3.27 V and $V_{7,10}$=-3.32 V). The capacitance of the island is $C_I = C_0 + C_g$, where the $C_0$ is the self-capacitance and $C_g$ the capacitance to the gates, which is predominantly the capacitance over the vacuum gap (because the dielectric constant of GaAs is 12x larger than vacuum). The plunger gate capacitance is ~$C_g$/3 (4 μm over 13 μm) by comparing the lengths of the gates. Using the capacitor model in b) we relate the change in capacitance to a change in separation $\Delta C_g/C_g = \Delta d/d$. The charge induced by change of the plunger gate $\Delta V_8$, $\Delta Q = C_g/3\cdot\Delta V_8$, should equal the charge induced by a change in chip separation, $\Delta Q = \Delta C\, V_g = C_g\, (\Delta d/d)\, V_g$. Thus an upper bound on the vibration amplitude is given by $\Delta d = d\, \Delta V_8/3V_g$ = 1.5 pm, using $V_g$ =-2.8 V, $d$ = 84 nm and $\Delta V_8$ = 0.15 mV.

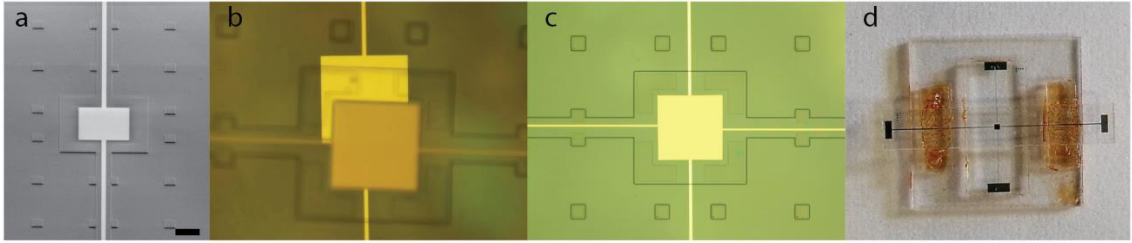

**Figure S2: Parallel plate capacitor.** (**a**) Tilted scanning electron micrograph of a quartz chip with a metallic square of 240 µm x 240 µm. The posts are ~100 nm higher than the metallic square and the surrounding quartz is etched 4 µm deep. The scale bar represents 100 µm. This quartz chip is flipped over and put onto the other quartz chip with the same metallic square but with neither posts nor etching. (**b**, **c**) Optical images of the two chips during and after alignment. (**d**) The full flip-chip assembly consisting of a parallel plate capacitor.

The two metallic squares act as a parallel plate capacitor from which the average distance is extracted by measuring the capacitance. The assembly is placed inside a nano-manipulator that is used to apply a variable force on the top of the assembly. Capacitance is measured using a lock-in amplifier. A sinusoidal voltage of 1 $V_{rms}$ is applied between the two plates at a frequency of 7.9 kHz. A measured current of 200 nA (Y-component) translates into 4 pF between the two capacitor plates resulting in an average separation of 120 nm. Precautions are taken to minimize the shunt capacitance of the wires to <0.2 pF.

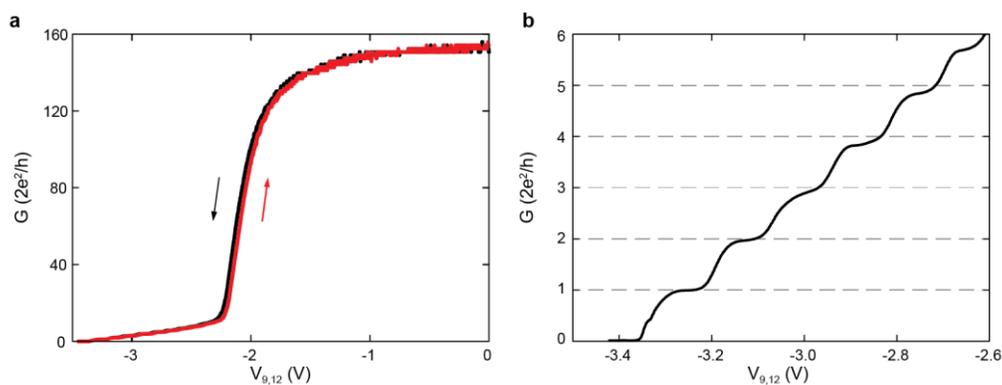

**Figure S3: Characterization of the right QPC for assembly A in the main text.** (**a**) Conductance as a function of the voltage on the right QPC (gates 9 and 12) in both sweeping directions. The electrons in the 2DEG underneath the gates are depleted at -2.2 V and pinch-off of the QPC occurs at -3.3 V. (**b**) Quantized conductance steps near pinch-off, similar to the left QPC shown in Fig. 2c and d in the main text.

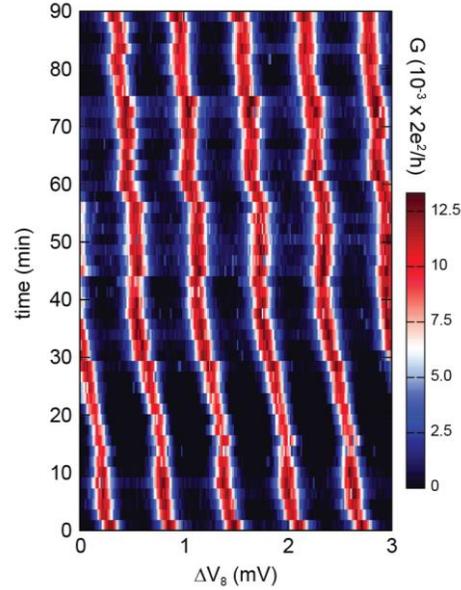

**Figure S4: Coulomb peaks as a function of time for assembly A in the main text.** By scanning the plunger gate voltage ($V_8$ = -2.5 V + $\Delta V_8$) over a 3 mV range, five Coulomb peaks of a quantum dot are probed at a voltage bias of $V_b$=10 µV. This trace is repeated 50 times during ~90 minutes. A slow drift in the Coulomb peak positions is observed. After 90 minutes, one extra electron has entered the quantum dot. Such drift may result from relaxation in the 2DEG and/or the setup. Note that except for the slow drift, no charge noise is observed, indicating no significant vibrations between the two chips. In the case that the drift is solely due to the mechanics of the setup, the gate-heterostructure separation needs to change by $\Delta d = d\, \Delta V_8/3V_g$ = 7.5 pm, using $\Delta V_8$=0.75 mV, $V_g$ =-2.8 V and $d$=84 nm.

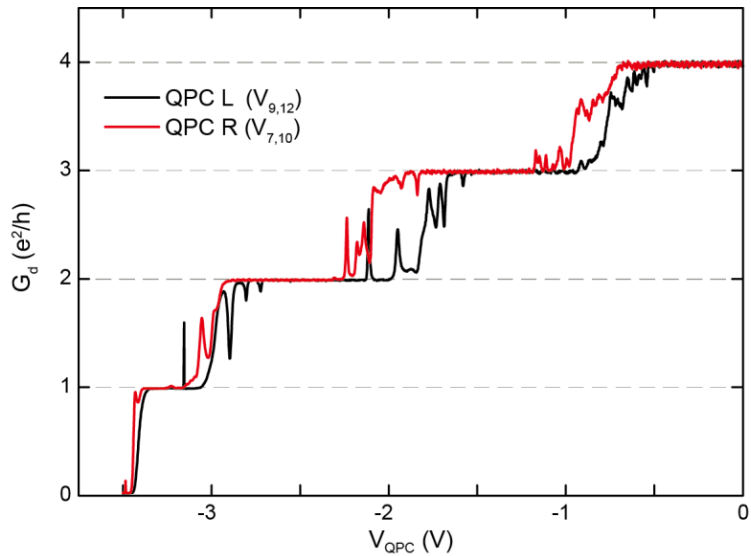

**Figure S5: QPC traces at a magnetic field of *B*=1.3 T for assembly A in the main text.** The figure shows the diagonal conductance through the left (black) and right (red) QPC as a function of the gate voltages $V_{7,10}$ and $V_{9,12}$, respectively, with the plunger gates set at the depletion point of -2.2 V. When $V_{7,10}$=0 or $V_{9,12}$=0, $G_d$=4$e^2$/h, consistent with a filling factor of v=4 in the bulk at *B*=1.3 T. Interference is observed if both QPC voltages are set on the low voltage side of plateau 3, $V_{7,10}$ = -1.6 V and $V_{9,12}$ = -1.9 V, as shown in Fig. 3c in the main text. At these gate settings the innermost edge channel (belonging to the fourth Landau level) is fully reflected at the QPC, while the third edge channel has a small backscattering amplitude. The outer two channels belonging to v=1 and v=2 are fully transmitted.

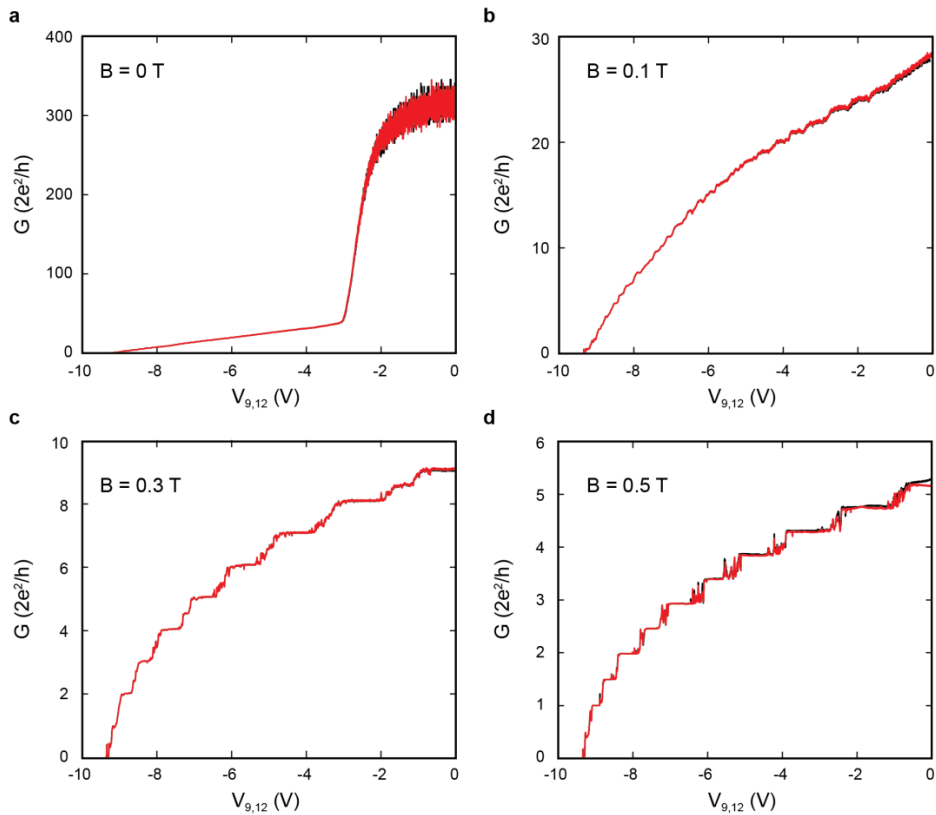

**Figure S6: QPC traces for assembly B in the main text.** (**a**, **b**, **c**, **d**) Diagonal conductance $G_d$ as a function of gate voltages $V_{9,12}$ for the right QPC of the interferometer at B=0, 0.1, 0.3, 0.5 T, respectively. At $V_{9,12}$=-3 V the electrons underneath the gates are depleted. Using a parallel plate capacitor model and the depletion voltage, a gate-heterostructure separation of ~115 nm is derived. At $V_{9,12}$=-9 V the QPC closes. However, due to the large QPC opening of 1.7 µm, quantized conductance is absent at zero magnetic field. In small perpendicular magnetic fields, quantized plateaux resulting from the backscattering of the quantum Hall edge channels are observed. Note that in all traces (a-b) no hysteresis is observed between down (black) and up (red) sweeps of the gate voltages.

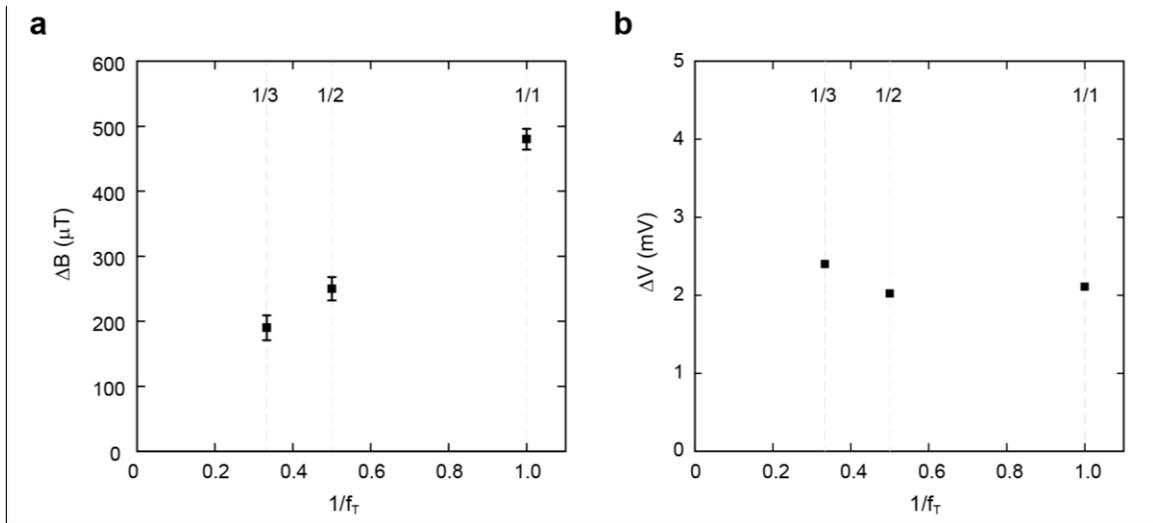

**Figure S7: Magnetic field and gate voltage period of quantum oscillations as a function of $1/f_T$ for Assembly B, where $f_T$ is the number of fully transmitted edge-channels.** (**a**) The magnetic field periods for different $f_T$ fall on a straight line through zero with a slope of 488.6 µT/($1/f_T$). (**b**) The period in the plunger gate voltage has no obvious dependence on $f_T$. The periods in (a) and (b) are determined from the interference patterns in Fig. 4 (b-g) in the main text after removing the background conductance, which is obtained by a $6^{th}$ order polynomial fit to the data. The change in the background conductance as a function of gate likely originates from details of the tunnel rates in the QPCs.

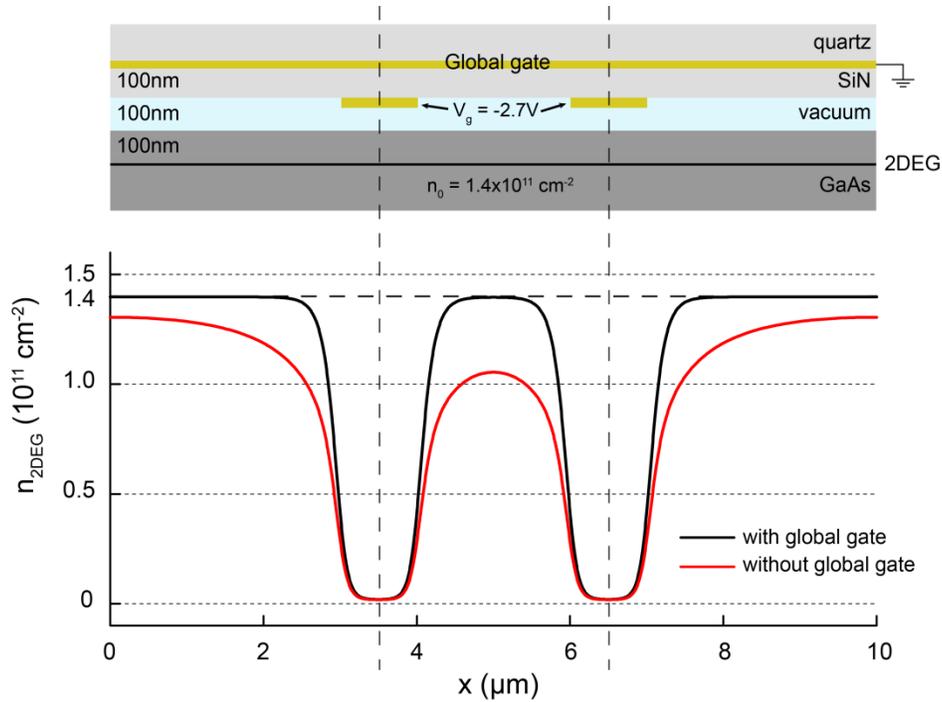

**Figure S8: Electrostatic potential simulation.** A double-layered gate structure can be implemented in the quartz gate-chip, as shown in the cross-section in the top panel. The electron density in the 2DEG can be simulated electrostatically for certain gate voltages. The red curve in the bottom figure shows the density profile in the 2DEG by applying a gate voltage of $V_g$=-2.7 V to the split gates (QPC gates in the interferometer) without implementing the global top gate. The black curve represents the density profile for a geometry with a grounded global top gate . The induced potential landscape is sharper in the latter case, and thus a better control can be achieved. In addition, as discussed in the main text, the top gate also increases the capacitance between gates and island (region enclosed by the interferometer) and can drive the interferometer from the Coulomb-dominated regime into the Aharonov-Bohm regime.

**Figure S9: Layout of the GaAs/AlGaAs heterostructure used in this experiment.**